\documentclass[11pt]{article}
\usepackage{a4wide}
\usepackage{amsfonts,amsmath,amssymb}
\usepackage{psfrag}
\usepackage{graphics}
\usepackage{subfigure}
\usepackage{ifpdf}
\usepackage{esint}
\usepackage{epsf}
\usepackage{color}
\ifpdf
\usepackage{epstopdf}
\usepackage{hyperref}
\else
\usepackage[hypertex]{hyperref}
\fi
\def\Xint#1{\mathchoice
{\XXint\displaystyle\textstyle{#1}}%
{\XXint\textstyle\scriptstyle{#1}}%
{\XXint\scriptstyle\scriptscriptstyle{#1}}%
{\XXint\scriptscriptstyle\scriptscriptstyle{#1}}%
\!\int}
\def\XXint#1#2#3{{\setbox0=\hbox{$#1{#2#3}{\int}$}
\vcenter{\hbox{$#2#3$}}\kern-.5\wd0}}

\def\dashint{\Xint-}
\renewcommand{\fint}{\dashint}
\def\ds{\displaystyle}

\newcommand{\vev}[1]{{\langle\;{#1}\;\rangle}} 
\newcommand{\eref}[1]{(\ref{#1})}

\newcommand{\blank}[1]{}

\newcommand\sect[1]{\section{#1}\setcounter{equation}0} 

\newcommand\void[1]       {}

\newcommand\be            {\begin{equation}}
\newcommand\bea           {\begin{eqnarray}}
\newcommand\rd             {{\mathrm d}}
\newcommand\ee            {\end{equation}}
\newcommand\eea           {\end{eqnarray}}

\def\Bone{{\hbox{1\!\!\!\!1}}}
\def\Bone{{\rm 1\!\!1}}

\newcommand\llangle       {\langle\!\langle}

\newcommand\one           {{\bf1}}

\newcommand\cG            {\mathcal{G}}

\newcommand\cH            {\mathcal{H}}

\newcommand\cO            {\mathcal{O}}

\renewcommand\vec[1]{{\vert{#1}\rangle}}

\newcommand\cev[1]{{\langle{#1}\vert}}
\newcommand\ccev[1]{{\llangle{#1}\vert}}

\newcommand\vac{{\vec{\, 0\,}}}
\newcommand\cav{{\cev{\, 0\,}}}

\def\3pt#1#2#3{{\langle{#1}|{#2}|{#3}\rangle}}

\def\cbI#1#2#3#4#5#6#7#8{
\setlength{\unitlength}{#1sp}%
\hbox to 2.8cm{\raise -2mm
\vbox{
\begin{picture}(1990,757)(4600,-2483)
\thicklines
{\put(4850,-1860){\line( 1,0){800}}}%
{\put(4850,-2460){\line( 1,0){800}}}%
{\put(5250,-1860){\line( 0,-1){600}}}%
\put(4775,-2460){\makebox(0,0)[rc]{$#2$}}
\put(4775,-1860){\makebox(0,0)[rc]{$#3$}}
\put(5750,-1860){\makebox(0,0)[lc]{$#4$}}
\put(5750,-2460){\makebox(0,0)[lc]{$#5$}}
\put(5300,-2160){\makebox(0,0)[lc]{$#6$}}
\end{picture}}}%
}

\def\cbb#1#2#3#4#5#6#7#8{
\setlength{\unitlength}{#1sp}%
\hbox to 3.6cm{\raise -2mm
\vbox{
\begin{picture}(2700,800)(4301,-2463)
\thicklines
{\put(4800,-1960){\line( 0,-1){500}}}%
{\put(5700,-1960){\line( 0,-1){500}}}%
{\put(4250,-2460){\line( 1, 0){2000}}}%
\put(4200,-2460){\makebox(0,0)[rc]{$#2$}}
\put(4800,-1900){\makebox(0,0)[cb]{$#3$}}
\put(4800,-2560){\makebox(0,0)[ct]{$#7$}}
\put(5250,-2380){\makebox(0,0)[cb]{$#4$}}
\put(5700,-1900){\makebox(0,0)[cb]{$#5$}}
\put(5700,-2560){\makebox(0,0)[ct]{$#8$}}
\put(6350,-2460){\makebox(0,0)[lc]{$#6$}}
\end{picture}}%
}}

\def\thefootnote{\fnsymbol{footnote}}
\begin{document}

\begin{flushright}  {~} \\[-12mm]
{\sf KCL-MTH-11-23}\\[1mm]
ITP-Budapest report No. 654
\\[1mm]
\end{flushright} 

\thispagestyle{empty}

\begin{center} \vskip 14mm
{\Large\bf   
Excited state g-functions from the Truncated
  Conformal Space}\\[20mm] 
{\large 
G.~Tak\'acs~~\footnote[1]{Email: takace@elte.hu} and G.M.T.~Watts~~\footnote[2]{Email: gerard.watts@kcl.ac.uk}
}
\\[8mm]
\footnotemark[1]
HAS Theoretical Physics Research Group,\\
1117 Budapest,
P\'azm\'any P\'eter s\'et\'any 1/A,
Hungary 
\\[8mm]
\footnotemark[2]
Department of Mathematics, King's College London,\\
Strand, London WC2R 2LS -- UK

\vskip 22mm
\end{center}

\begin{quote}{\bf Abstract}\\[1mm]
In this paper we consider excited state g--functions, that
is, overlaps between boundary states and excited states in boundary
conformal field theory. 
We find a new method to calculate these overlaps numerically using a
variation of the truncated conformal space approach.
We apply this method to the Lee-Yang model for which the unique boundary
perturbation is integrable and for which the TBA system describing the
boundary overlaps is known.
Using the truncated conformal space approach we obtain numerical
results for the ground state and the first three excited states which
are in excellent agreement with the TBA results.
As a special case we can calculate the standard g-function which is
the overlap with the ground state and find that our new method is
considerably more accurate than the original method employed by Dorey
et al. 

\end{quote}


\vfill
\newpage 

\setcounter{footnote}{0}
\def\thefootnote{\arabic{footnote}}

\sect{Introduction}

If a conformal field theory is defined on a domain with a boundary
then boundary conditions need to be defined. In the simplest case of
the unit disk, the boundary conditions are implemented through a
boundary state. 
Such boundary states are not normalisable, but are uniquely
determined by their overlaps with the normalisable, finite-energy
states of the bulk system.
The exact forms of the boundary states of all the 
conformal boundary conditions for many conformal field theories, and
hence all these overlaps, have
been known for some long time, 

In this paper we consider non-conformal boundary conditions generated
by a perturbation of a conformal boundary condition by the integral of
a local field along the boundary. 
The resulting perturbed boundary condition can also be described by a
boundary state which is uniquely specified by 
its overlaps with the bulk states.
If the perturbation is integrable then it may be possible to calculate
the overlaps using the Thermodynamic Bethe Ansatz (TBA). 
For such systems there is a distinguished basis of states and for
each state there is an integral equation for the overlap.
If the perturbation is not integrable then one needs an alternative
method to calculate the overlaps and determine the boundary state.

One important numerical method to study perturbed conformal field
theory is the Truncated Conformal Space Approach (TCSA). This was
initiated by Yurov and Zamolodchikov \cite{YZ1} and used to study the
finite-size spectrum of bulk perturbations of conformal field
theories.  It was adapted by Dorey et al to study the finite-size
spectrum of boundary-perturbed conformal field theory
\cite{Dorey:1997yg}, and it was shown by Dorey et al in
\cite{Dorey:2000} how this could be used to calculate the g--function,
or ground state overlap with the boundary state.
It was not possible, however, to use this method or a modification to
calculate overlaps of the boundary state with excited bulk states.

Here, we present an alternative method based on the TCSA
inspired by the work of Zamolodchikov on the partition function of a
perturbed conformal field theory on a sphere \cite{Zamo:sphere} where
the partition function on a sphere was calculated using the
Schr\"odinger equation for a truncated system. The sphere was
conformally mapped into the complex plane and the resulting system was
formally the same as the usual truncated conformal space approach with
a space-dependent coupling.

We calculate the boundary state overlaps as one point
functions of bulk fields on a disk. The disk can be conformally mapped
to the upper half plane and the one-point function of a bulk field is
mapped to the expectation value of the bulk field in a state
corresponding to a half-disk with a time-dependent coupling to a
boundary field. 
We then truncate the Hilbert space and numerically integrate the
perturbation to find the state for the half-disk.
Once the half-disk state has been found, it is only necessary to
calculate the matrix elements of the required bulk field to find the
overlap of the bulk state with the boundary state.

There are some technical difficulties applying this method to higher
excited states arising from the form of the TCSA expression for the
bulk field expectation value which  is similar to a Fourier expansion
in the argument of the insertion point. Excited states correspond to
repeated derivatives of the field and the derivatives of the Fourier
series do not themselves converge. this can be handled by 
forcing convergence of the Fourier series (at the loss of some
information) which gives numerically satisfactory results at the
expense of some complication in the calculational scheme.

We apply this method to the simplest perturbed boundary conformal
field theory, the Lee-Yang model, which was the model studied in
\cite{Dorey:1997yg,Dorey:2000} and has the advantage that the
overlaps can also be calculated using the TBA method and we find
excellent agreement.

The layout of the paper is as follows.  In section 2 we outline the
general truncated space method and in section 3 we apply it to the
ground state and first excited state in the Lee-Yang model. In section
4 we consider the next two excited states and explain how to force
convergence of method. We outline the TBA method used in an appendix.

\sect{$g$-functions and perturbed conformal field theory}
\label{sec:disk}

Our aim is to calculate the overlap of a bulk state with the boundary
state corresponding to a perturbation of a conformal boundary
condition  
on the boundary of the disc $|z|=R$ 
by the insertion of 
$\exp(-\lambda \int \psi(x) \rd x)$ on this boundary. We take $\psi$
to be a primary boundary field of conformal dimension $h$. 
We shall denote such an overlap by $\cG_i$ where $i$ labels the
particular bulk state.
The most natural way to calculate this one point function is to expand
the exponential and consider it as a perturbative expansion with
repeated insertions of the boundary field.
If $h<1/2$, then 
such a function $\cG_i$ will have a regular expansion in powers of 
$(\lambda R^y)$ for small 
$\lambda$ (where $y=1-h$ is the mass dimension of the coupling
constant $\lambda$), but for large $\lambda$ the dominant term is 
non-perturbative:
\be
 \log \cG_i
\sim - \tilde R f_B + \ldots
\;,
\ee
where $f_B$ is the boundary free energy per unit length and 
$\tilde R = 2\pi R$
is the length of the boundary.
While this is the natural consequence of a perturbative expansion, it
is many ways more natural to consider the state perturbed boundary to
interpolate between two conformal boundary conditions which requires
subtracting the free energy term; this gives the definition of the
(non-perturbative) $g$-function $g_i$ as 
\be
  g_i(\lambda R^y) 
= \exp(2\pi f_B R)\, \cG_i(\lambda R^y)
\;.
\ee
We shall define the perturbed boundary state 
$\cev{B(\lambda)}$ 
by the conformal normalisation so that the overlap gives with the bulk
state $\vec{\varphi_i}$ the
perturbative $\cG$-function,
\be
  \cG_i(\lambda R^y) 
= \cev{B(\lambda)}\,\varphi_i\, \rangle
\;.
\ee
The overlap 
$\cev{B(\lambda)}\varphi_i\rangle$ 
can be realised as the insertion of the field $\varphi_i(0)$ in the 
disk with boundary condition $B(\lambda)$, where
$\vec\varphi_i = \varphi_i(0)\vac$
and it is this definition which is the basis of the TCSA method
described below.

\subsection{The TCSA approach to the disk}
\label{ssec:op}

In order to use the TCSA method, we need to find a way in which 
the boundary field can be constructed as an operator acting on a
Hilbert space (which can be truncated), or equivalently in which the
system has a Hamiltonian description where the boundary field enters
the Hamiltonian. 
We first outline the operator realisation and then the Hamiltonian
description.

\begin{figure}
\centering{$  
 \begin{picture}(380,84)
  \put(0,0){\scalebox{0.2}{\includegraphics{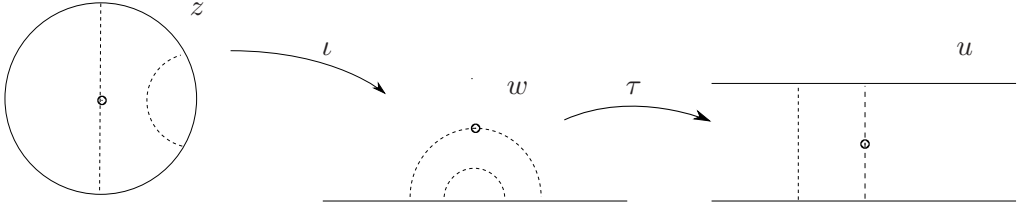}}}
  \put(0,0){
     \setlength{\unitlength}{1pt}\put(0,0){
     \put(70,80)   {$z$}
     \put(120,65)   {$\iota$}
     \put(190,50)   {$w$}
     \put(235,50)   {$\tau$}
     \put(360,65)   {$u$}
     }\setlength{\unitlength}{1pt}}
  \end{picture}
$}
\caption{The maps $\iota:z\mapsto w$ and $\tau:w \mapsto u$, showing
  the ``equal-time'' quantisation surfaces as dashed lines.}
\label{fig:drawing}
\end{figure}

To have an operator action, once must choose ``equal-time'' surfaces
to which one can associate a Hilbert space of states on these
surfaces. 
For the boundary field to act, these surfaces must have boundaries,
and the most natural surface is a line segment corresponding to equal
time slices across an infinite strip. This is conformally equivalent
to the upper half plane with the equal-time surfaces being semi-circles
of constant radius, known as radial quantisation.
We can map the interior of the unit disk (with coordinate $z$) to the
upper half plane (with coordinate $w$) using the map
\be
\iota: z \mapsto w = -i \frac{z-1}{z+1}
\;.
\label{eq:iota}
\ee
This map is shown in figure \ref{fig:drawing}
The one-point function in the perturbed boundary condition $\langle
\varphi(0) \rangle_{\lambda}$ becomes the matrix element of a field
between states representing the interior of the unit semi-circle (with
perturbed boundary condition) and the exterior (with perturbed
boundary condition).
Using the transformation property of primary spinless boundary fields,
\be
 \iota: \phi(z) \mapsto \left| \frac{\rd w}{\rd z} \right|^h \phi(w)
\;,\;\;
  \phi(\exp(i\,\theta)) \mapsto \left(\frac{1+t^2}{2}\right)^{h} \phi(t)\;,\;\;
    t = \tan(\theta/2)
\;,
\label{eq:phimap}
\ee
we can write the state corresponding to the interior of the 
semi-circle of radius $r$ as
\be
\vec{\Psi(r)}
 = P \exp\Big( - \lambda \int_{t=0}^r 
   (\phi(t) + \phi(-t))
   (\tfrac{2}{1+t^2})^y \,\rd t \Big)\,\vac
\;,
\ee
where $y=1-h$ and $P$ denotes radial ordering. 
The overlap $\cev{B(\lambda)} \varphi \rangle$ is then
\be
  \cev{B(\lambda)} \varphi \rangle
= \hbox{const.}\,
\cev{\Psi(1)} \imath(\varphi(0)) \vec{\Psi(1)}
\;,
\label{eq:1}
\ee
where (const.) is a constant normalisation factor and
where $\iota(\varphi(0))$ is the image of the field $\varphi(0)$ under the map
$\iota$; if $\varphi(z,\bar z)$ is a spinless quasi-primary field 
of total conformal weight $\Delta$, this is the simple rescaling
\be
  \iota(\varphi(z,\bar z)) 
= \left| \frac{\rd w}{\rd z}\right|^\Delta\,
  \varphi(w,\bar w)
\;,
\ee
otherwise it is more complicated. 

The state $\vec{\Psi(r)}$ satisfies the differential equation
\be
\frac{\rd}{\rd r}\vec{\Psi(r)}
= 
 - \lambda \big(\phi(r) + \phi(-r)\big)
   (\tfrac{2}{1+r^2})^y \,\vec{\Psi(r)}
\;.
\label{eq:op1}
\ee
and 
we can use \eref{eq:op1} as the basis for a TCSA calculation.
We can truncate the Hilbert space of the system at excitation level
$N$ and solve the first order matrix differential equation for
$\vec{\Psi(r)}_N$ using the 
finite matrix truncations of $(\phi(r)+\phi(-r))$.
We can then take the matrix elements of $\iota(\varphi)$ in the
truncated state $\vec{\Psi(1)}_N$ and this is the TCSA approximation
to the overlap.

\subsection{Realisation as a Hamiltonian system}

The calculation of the previous section can be recast in terms of a
perturbed Hamiltonian (in the spirit of \cite{Zamo:sphere}) as
follows.
It is easiest to see the relation  by mapping
the disk to an infinite strip, by the further map
\be
\tau:  w \mapsto u = \log(w)
\;,
\ee
which maps the original disk to the strip
$0\leq\mathrm{Im}(u)\leq\pi$ (see figure \ref{fig:drawing}.
If $u=x+iy$ then 
the lines $x=\mathrm{const.}$ can be taken as equal time slices and
the Hamiltonian propagates in $x$ along the strip. 
The equation \eref{eq:op1} is transformed into the Schr\"odinger equation
in the interaction picture,
\be
\frac{\rd}{\rd x}\vec{\Psi(x)}
= - H_I(x)\,\vec{\Psi(x)}
\;,
\label{eq:ham1}
\ee
where $H_I(x)$ is the $x$-dependent Hamiltonian
\be
  H_I(x)
= \lambda\, \mathrm{sech}(x)^y (\phi(x) + \phi(x + i\pi))
\;.
\ee
This problem can be also be formulated in the Schr\"odinger picture in which
case the relevant Hamiltonian is
\be
  H(x) 
= L_0 + \lambda\, \mathrm{sech}(x)^y (\phi(0) + \phi(i\pi))
\;.
\ee
It is this formulation which is directly comparable to the Hamiltonian
formulation of the bulk model on the sphere in \cite{Zamo:sphere}. 
In practice we have used \eref{eq:op1}
to perform our TCSA
calculations as it is numerically more stable.

We now turn to the calculation of the ground state and first excited
state overlaps in the Lee-Yang model.

\sect{The Lee-Yang model}
\label{sec:YL1}

The Lee-Yang model has been used extensively as a test of exact and
numerical methods because it is in many ways the simplest, and the 
convergence of the TCSA calculation of its spectrum is amongst fastest
of all known models. 
In this context it is the simplest because it has only two conformal
boundary conditions and a single integrable perturbation which
connects them. It also has the fewest states at any particular
excitation level so that the size of any TCSA truncation will be
smallest in the Lee-Yang model.
The details of the conformal and integrable descriptions of the
boundary conditions can be found in \cite{Dorey:1997yg,Dorey:2000} and
we only recall the most important details here.

We start with the description of the theory in the bulk.
The Lee-Yang model is the minimal conformal field theory with
$c=-22/5$. It is non-unitary and there are two highest weight
representations of the Virasoro algebra which are relevant, with
conformal weights $-1/5$ and $0$. 
If we denote these two highest weight representations by $H_{-1/5}$
and $H_0$ then the Hilbert space of the bulk theory is
\be
  \cH_{bulk} 
= 
(H_0 \otimes \bar H_0) 
\; \oplus \;
(H_{-1/5} \otimes \bar H_{-1/5}) 
\;.
\label{eq:Hbulk}
\ee
We take the bulk highest weight states to be $\vac$ and $\vec\varphi$ of
conformal weight $0$ and $-2/5$ respectively. The $SL(2)$ invariant
state is $\vac$ but the ground state in the bulk theory is
$\vec{\varphi}$. It is convenient to
normalise these states as
\be
  \cev{\, 0\,}\, 0\, \rangle=-1
\;,\;\;
  \cev \varphi \varphi \rangle= 1\;.
\ee
The state $\vec\varphi$ is associated to the primary field
$\varphi(z,\bar z)$ and satisfies
\be
  \vec\varphi = \varphi(0,0)\,\vac
\;.
\ee
The Lee-Yang model has two conformal boundary conditions. We will
adopt the conventions of \cite{Dorey:1997yg} in which these are
called 
$\Bone$ and $\Phi$ and 
the states associated to them on the unit
disk are
\be
\cev{B_\Phi} = g_\Phi \ccev\varphi - Z_\Phi \ccev 0
\;,\;\;
\cev{B_\Bone} = g_\Bone \ccev\varphi - Z_\Bone \ccev 0
\;,\;\;
\label{eq:bss}
\ee
where $g_\alpha$ is the $g$-value of the boundary condition $\alpha$,
$Z_\alpha$ is the value of the disk partition function with 
boundary condition $\alpha$ and $\ccev h$ are the usual Ishibashi
states \cite{ishibashi}.

As described in \cite{Dorey:1997yg}, there is a single integrable flow
between the boundary conditions $\Phi$ and $\Bone$, the 
perturbed boundary conditions being denoted by $\Phi(h)$ where $h$ is
the coupling to the unique relevant boundary field on the condition
$\Phi$. This field has conformal weight $-1/5$ and is denoted by $\phi$.

The particular boundary conditions we are considering, both perturbed
and unperturbed, are 
rotationally invariant and so 
the only states which have non-zero overlap must have zero spin, that
is they must have equal $L_0$ and $\bar L_0$ eigenvalues.
The lowest lying zero-spin states, their conformal weights and the fields to
which they correspond are in table \ref{tab:lystates}. We'll denote
the overlap with the $i$-th excited state as $\cG_i$ and the
associated $g$-function as $g_i$. It is a property of the conformal boundary
states that the overlap with a normalised bulk state only depends on the
representation in which it lies, so that in particular we have
\be
  \cG_0(0) = \cG_2(0)=\cG_3(0)=\cG_5(0) 
= \cev{B_\Phi}\,\varphi\,\rangle=g_\Phi
\;,\;\;
  \cG_1(0) = \cG_4(0) 
= \cev {B_\Phi}\,0\,\rangle=-Z_\Phi
\;.
\ee
\begin{table}[bht]
{\renewcommand{\arraystretch}{1.4}
$$\begin{array}{r@{\,}lcr@{\,}l}
&\hbox{state}&\hbox{weight}&&\hbox{field}\\ \hline
&\vec{\varphi} & -2/5 && \varphi(z) \\
&\vac          &  0   && \Bone \\
-\tfrac 52
&L_{-1}\bar L_{-1}\vec\varphi
              & 8/5  &
-\tfrac 52
& \partial\bar\partial\varphi(z,\bar z)
\\
-\tfrac{25}{12}
&L_{-1}L_{-1}\bar L_{-1}\bar L_{-1}\vec\varphi
              & 18/5  &
-\tfrac{25}{12}
& \partial^2\bar\partial^2\varphi(z,\bar z)
\\
-\tfrac{11}{5}
&L_{-2}\bar L_{-2}\vac & 4 &
-\tfrac{11}{5}
& T (z) \bar T (\bar z) \\
-\tfrac{125}{288}
&L_{-1}L_{-1}L_{-1}\bar L_{-1}\bar L_{-1}\bar L_{-1}\vec\varphi
              & 28/5  &
-\tfrac{125}{288}
& \partial^3\bar\partial^3\varphi(z,\bar z)
\\
\end{array}
$$
\caption{Lowest lying spin zero states in the Lee-Yang model}
\label{tab:lystates}
}
\end{table}

We also need the description of the $\Phi$ boundary condition in the
upper-half plane operator picture.
In this formulation, the unperturbed
Hilbert space is that of a strip with the $\Phi$ boundary condition on
both sides
and
is the direct sum of the two Virasoro
algebra representations in the Lee-Yang model,
\be
\cH_{strip} = 
H_0 \oplus H_{-1/5}
\;.
\label{eq:Hstrip}
\ee
We denote the highest weight states in the UHP picture as $\vac$ and
$\vec\phi$ respectively. 
The ground state  is $\vec{\phi}$ which
corresponds to the insertion of the boundary field $\phi(0)$ at the
origin of the model on the upper half plane. 
We again choose to normalise these states as
\be
  \cav \,0\,\rangle = -1
\;,\;\;
  \cev\phi \phi \rangle = 1
\;.
\ee
The expectation value of any operator $\cO$ on the disk with the
conformal $\Phi$ boundary 
condition is now given by
\be
\vev{\cO}_\Phi = - Z_\Phi \cav \,\iota(\cO)\, \vac
\;,
\label{eq:expect}
\ee
where $\iota$ is the map \eref{eq:iota} from the disk to the UHP.
In particular we recover the partition function $Z_\Phi$ for the
identity operator $\cO=1$,
\be
{\vev 1}_\Phi = - Z_\Phi \cav\,0\,\rangle = Z_\Phi
\;.
\ee
For the ground state entropy we use
the operator $\cO = \varphi(0,0)$ and $\iota(\varphi(0,0)
= 2^{-2/5}\varphi(i,-i)$.
With the result
$
  \cav \varphi(w,\bar w)\vac 
= - {}^{(\Phi)}\!B_\varphi^\Bone
  (2r\sin\theta)^{2/5}
$
we get
\be
  \vev{ \varphi(0) }_\Phi
= - Z_\Phi \cav\, 2^{-2/5}\varphi(i,-i)\,\vac
= Z_\Phi\, {}^{(\Phi)}\!B_\varphi^\Bone
= g_\Phi
\;.
\ee
We now turn to the overlaps with the excited boundary states.

The first overlap studied was the ground state overlap, the original
$g$-function of \cite{AffleckLudwig} for which a TBA formulation was
first described in \cite{LMSS}. Subsequently, it was shown in
\cite{BLZ:9607099} that the $g$-function was the first of an infinite
set of excited state $g$-functions which arose as the eigenvalues of
generalised transfer matrices satisfying a so-called $T$-system. This
$T$-system has been put into an orthodox conformal field theory
context by Runkel \cite{Runkel-T}. In the case of the Lee-Yang model
the $T$-system is identical to the so-called $Y$-system and Bazhanov et
al also showed how to find the TBA system for excited state
$Y$-functions and solved it for the first excited state.
The ground state $Y$-function was compared to an alternative TCSA
calculation of $\cG_0$ in \cite{Dorey:2000} with reasonable agreement.
Here we state the expected relation between the $\cG$ and $Y$
functions:
\bea
    Y_0(\theta)
 =  \frac{y_0}{g_\Phi}
    \cev{B_\Phi(h)} \,\varphi\, \rangle
\;,\;\;
    Y_1(\theta)
 =  -\frac{y_1}{Z_\Phi}
    \cev{B_\Phi(h)} \,0\, \rangle
\;,
\label{eq:GY}
\eea
where 
$y_0 = \tfrac 12(1+\sqrt 5)$,
$y_1 = \tfrac 12(1-\sqrt 5)$ and 
we have suppressed the $h$-dependence of the state
$\vec{\Psi(1)}$.
The $Y$-functions depend on a rapidity $\theta$
which is related to $h$ by
\cite{BLZ:9607099}
\be
   h (2\pi)^{6/5} 
= - \tfrac 12 \,h_c \,e^{6 \theta/5}
\;,\;\;
h_c =
- \pi^{3/5}
  \, 
  \frac{ 2^{4/5}
  \, 5^{1/4}\,\sin \tfrac{2\pi} 5 }
       { ( \Gamma( \tfrac 35 ) \Gamma(\tfrac 45) )^{1/2} }
  \left( 
  \frac{ \Gamma( \tfrac 23 ) }{\Gamma( \tfrac 16 )}
  \right)^{\!\!6/5}
\!\!= -0.68529...
\ee
The extra factor of $(2\pi)^{6/5}$ between the formula here and that
in \cite{Dorey:2000} arises because the boundary of the
unit disk has length $2\pi$ and it is the combination $h\tilde R^{6/5}$
which is relevant to this equation.


We now describe the TCSA approximation and in particular the lowest
level truncation where the UHP
Hilbert space is restricted to just two states.

\subsection{TCSA method and results}

We have used the TCSA approximation to the differential equation
\eref{eq:ham1} to calculate approximations to the state
$\vec{\Psi(1)}$ and then calculated the
exact expectation value of the fields $\varphi(i)$ and $\Bone$ in
these approximate states for truncation levels 0 up to 20, that is for
systems of size 2 up to 323. 
To illustrate the method, we consider here the case of truncation to
level 0, that is we truncate the 
Hilbert space to just the two highest weight states $\vac$ and
$\vec\phi$, and construct 
the TCSA equations for the upper half plane state \eref{eq:op1}
explicitly.

\begin{figure}[htb]
\centering{$  
 \begin{picture}(330,190)
  \put(0,0){\scalebox{1.0}{\includegraphics{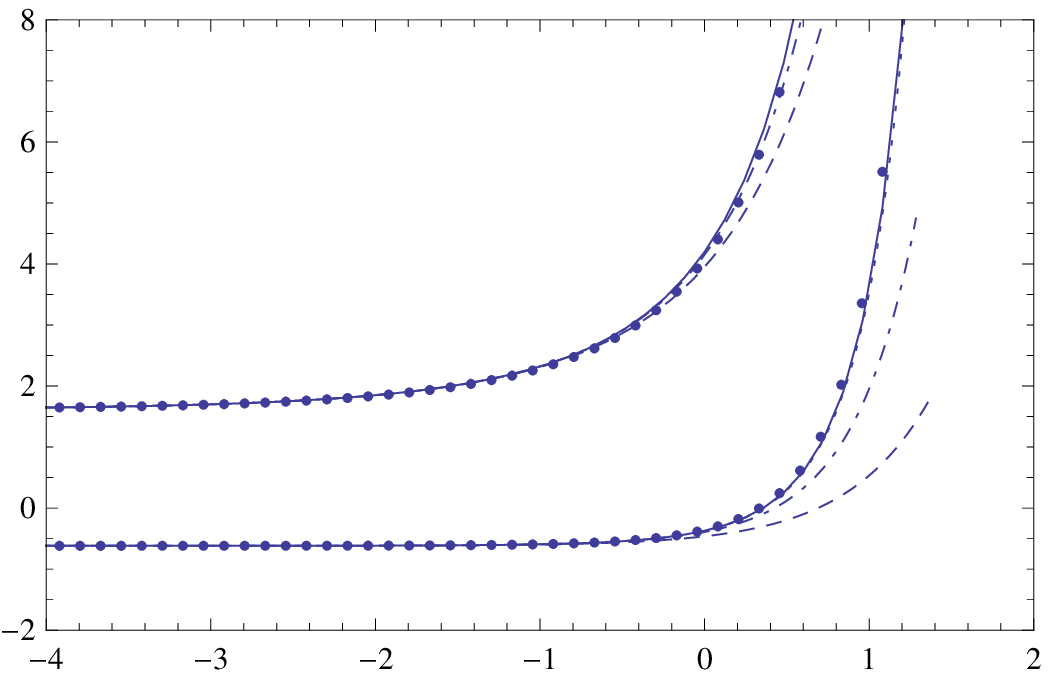}}}
  \put(0,0){
     \setlength{\unitlength}{.80pt}\put(0,0){
     \put(220,32)   {$Y_1$}
     \put(190,92)   {$Y_0$}
     }\setlength{\unitlength}{1pt}}
  \end{picture}
$}

\caption{The two lowest lying $Y$-functions in the Lee-Yang model, as
  calculated from TBA 
  equations (solid lines), and the TCSA approximations calculated using
  the space truncated to 2 states (points) plotted against
  $\log(h(2\pi)^{6/5})$ for comparison with \cite{BLZ:9607099}. Also
  shown are the second order perturbation theory expressions (dashed
  lines), third order (dot-dashed) and sixth order (dotted) -- 
see appendix \ref{sec:pt} for details. } 
\label{fig:level0}
\end{figure}

We take the state $\vec{\Psi(r)}$ to be truncated to just the highest
weight states $\vec\phi$ and $\vac$, so that it is determined by just
two functions $f_\phi(r)$ and $f_0(r)$,
\be
  \vec{\Psi(r)} 
= f_\phi(r)\, \vec\phi \;+\; f_0(r)\, \vac
\;.
\ee
The actions of the field $\phi(r)$ 
on these states are (for $r$ both positive and negative)
\bea
    \phi(r)\, \vac 
&=& \vec\phi + r L_{-1}\vec\phi +  \ldots
\;,\;\; 
\\
  \phi(r)\,\vec{\,\phi\,}
&=& - |r|^{2/5}\left(\,\vac + \tfrac 1{11} r^2 L_{-2}\vac + \ldots\right)
\nonumber\\&& 
\;\;  + \; c_{\phi\phi}{}^\phi |r|^{1/5}
    \left(\,\vec\phi + \tfrac 12 r L_{-1}\vec\phi +  \ldots\right) 
\;.
\eea
This means that truncating \eref{eq:op1} to the two states gives the
differential equation
\be
\frac{\rd}{\rd r}
\begin{pmatrix}
f_0(r) \\
f_\phi(r) 
\end{pmatrix}
= - h
  \, \left(\tfrac{2}{1+r^2}\right)^y
  \, 
\begin{pmatrix}
 0 & - 2 r^{2/5} \\
 2 & 2 c_{\phi\phi}{}^\phi r^{1/5}
\end{pmatrix}
\,
\begin{pmatrix}
f_0(r) \\
f_\phi(r) 
\end{pmatrix}
\;,
\ee
which should be solved subject to the initial conditions
\be
f_0(0) = 1
\;,\;\;\;\;
f_\phi(0) = 0
\;.
\ee
The result is the TCSA estimate
\be
 \vec{\Psi(1)}
= a \vac \;+\; b \vec{\,\phi\,}
\;,\;\;\;\;
a = f_0(1)\;,\;\;b=f_1(1)
\;.
\label{eq:Psiform}
\ee
From this we can calculate the TCSA estimates of the two $\cG$-functions,
\bea
 \cG_0
&=& - Z_\Phi\,2^{-2/5}\,
    \cev{\Psi(1)}\,\varphi(i,-i)\,\vec{\Psi(1)}
\;,\;\;
\\[1mm]
\cG_1
&=& - Z_\Phi\,
    \cev{\Psi(1)} {\Psi(1)} \rangle
\;.
\eea
To complete the calculation of the $Y$-functions, we need the one-point functions of the
field $\varphi(w,\bar w)$ on the UHP, which are in appendix
\ref{app:ly} and the re-scalings from \eref{eq:GY}, resulting in
\bea
 Y_0
&=& y_0 
\Big( a^2 
\;-\;
 2^{4/5} \frac{{}^{(\Phi)}\!B_\varphi^\phi}{{}^{(\Phi)}\!B_\varphi^1}\, a b
\;+\;
 2^{-9/5}\frac{(\sqrt 5 - 1)\,\Gamma(\tfrac 1{10})\Gamma(\tfrac 15)}
      {5\sqrt\pi\,\Gamma(\tfrac 45)}\, b^2
\Big)
\nonumber\\
&=&
 1.618034..(\, a^2 \; - \; 2.50895..\, a b 
 + 1.50258..\, b^2)
\;,\;\;
\\[1mm]
 Y_1
&=& y_1\, (a^2 - b^2)
\;=\; 
    - 0.618034..\,(a^2 - b^2)
\;.
\eea
We used Mathematica \cite{mathematica}
built-in routines to solve the matrix differential equations with
satisfactory results.
Already the accuracy obtained using just
a truncation to 2 states is quite surprising.
We list some numerical results in table \ref{tab:level0} and 
in figure
\ref{fig:level0} we plot the 
2-state truncation estimates for
the functions $Y_0$ and $Y_1$ together with the exact results
calculated using the TBA equations of \cite{BLZ:9607099} and the
results of second and third order perturbation theory (see appendix
\ref{sec:pt}).  
\blank{ 
\begin{table}
\[
\begin{array}{l|ll|ll|ll}
\log h & a & b & G_0^{TCSA} & G_0^{exact} & G_1^{TCSA} & G_1^{exact}
\\
\hline
 -\infty & \phantom{-}1 & \phantom{-}0 
& \multicolumn{2}{c|}{ 1.17319 }
& \multicolumn{2}{c}{ -0.570017 }
\\
-5 & \phantom{-}0.999783  & -0.0237672
& \phantom{2}1.24362  & \phantom{2}1.24675
& -0.569448 & -0.569463
\\
-4 & \phantom{-}0.998362 & -0.0669088
& \phantom{2}1.37387  & \phantom{2}1.38285
& -0.565599 & -0.565732
\\
-3 & \phantom{-}0.987093 & -0.200034
& \phantom{2}1.79484  & \phantom{2}1.82318
& -0.532589 & -0.533920
\\
-2 & \phantom{-}0.886238 & -0.704015
& \phantom{2}3.63168  &  \phantom{2}3.74303
& -0.165179 & -0.186155
\\
-1 & -0.379141 & -3.85251
& 22.0327  & 22.5198
& \phantom{-}8.37817 & \phantom{-}7.41159
\end{array}
\]
\caption{Numerical results for the two-state TCSA approximation}
\label{tab:level0}
\end{table}
}

\begin{table}[bht]
\[
\begin{array}{l|ll|ll|ll}
\log h &  \phantom{-}a & \phantom{-}b & \phantom{-}Y_0^{TCSA} & \phantom{-}Y_0^{exact} & \phantom{-}Y_1^{TCSA} & \phantom{-}Y_1^{exact}
\\
\hline
 -\infty & \phantom{-}1 & \phantom{-}0 
& \multicolumn{2}{c|}{ 1.61803 }
& \multicolumn{2}{c}{ -0.618034 }
\\
-6 & \phantom{-}0.999783  & -0.0237672
& \phantom{2}1.65316  & \phantom{2}1.65471
& -0.617952 & -0.617953
\\
-5 & \phantom{-}0.999783  & -0.0237672
& \phantom{2}1.71517  & \phantom{2}1.71948
& -0.617417 & -0.617433
\\
-4 & \phantom{-}0.998362 & -0.0669088
& \phantom{2}1.89480  & \phantom{2}1.90719
& -0.613244 & -0.613387
\\
-3 & \phantom{-}0.987093 & -0.200034
& \phantom{2}2.47539  & \phantom{2}2.51448
& -0.577453 & -0.578895
\\
-2 & \phantom{-}0.886238 & -0.704015
& \phantom{2}5.00870  &  \phantom{2}5.16228
& -0.179094 & -0.201836
\\
-1 & -0.379141 & -3.85251
& 30.3869  & 31.0587
& \phantom{-}9.08392 & \phantom{-}8.03592
\end{array}
\]
\caption{Numerical results for the two-state TCSA approximation}
\label{tab:level0}
\end{table}

\begin{table}[bht]
\[
\begin{array}{c@{}rllllll}
& N & g_0 & g_1 & g_2 & g_3 & \phantom{-}g_4 & \phantom{-1}g_5 \\ \hline
& 0 & 0.853057 & 0.0387796 & 0.430042 & 0.561775 &\phantom{-}0.106314&  \phantom{-1}0.687637 \\
& 2 & 0.889914 & 0.0422042 & 0.455962 & 0.653033 &\phantom{-}0.103007&  \phantom{-1}0.993606 \\
& 4 & 0.878654 & 0.0428922 & 0.437669 & 0.487325 &\phantom{-}0.125051&  \phantom{-1}0.688144 \\
& 6 & 0.882040 & 0.0431683 & 0.451097 & 0.745018 &\phantom{-}0.076786&  \phantom{-1}2.72391 \\
& 8 & 0.879529 & 0.0433135 & 0.440139 & 0.400679 &\phantom{-}0.153111&  \hbox to 0.7pt{}-0.385525 \\
&10 & 0.880638 & 0.0434018 & 0.449176 & 0.830633 &\phantom{-}0.048286&  \phantom{-1}6.80768 \\
&12 & 0.879593 & 0.0434606 & 0.441394 & 0.322728 &\phantom{-}0.181909&  \hbox to 0.7pt{}-3.52987 \\
&14 & 0.880125 & 0.0435024 & 0.448145 & 0.907610 &\phantom{-}0.019363&  \phantom{-}14.0118\\
&16 & 0.879565 & 0.0435335 & 0.442140 & 0.251577 &\phantom{-}0.210946&  \hbox to 0.7pt{}-9.55319 \\
&18 & 0.879871 & 0.0435574 & 0.447506 & 0.978097 &-0.009753          &  \phantom{-}25.0185\\
&20 & 0.879526 & 0.0435764 & 0.442631 & 0.185561 &\phantom{-}0.240248&  -19.1595\\
\hline
&\infty^{(P)}   & 0.8794   & 0.04368   & 0.4449   & 0.6020   &\phantom{-}0.1079  & \hbox to 0.7pt{}-6.93 \\
&\infty^{(Q)}
   & 0.8794   & 0.04358   & 0.4449   & 0.6143   &\phantom{-}0.1008  & \phantom{-1}0.6912 \\
\hline
\multicolumn{2}{r}{\hbox{TBA} }
   & 0.879214 & 0.0437266 & 0.444889 & 0.614126 &\phantom{-}0.100843& \phantom{-1}0.690846 
\end{array}
\]
\caption{The values of $g_0, g_1, g_2, g_3, g_4$ and $g_5$ for $\log h = -2 $ 
estimated from TCSA at
  truncation levels $0\leq N\leq 20$, the values obtained by Pad\'e
  extrapolation of the series (P) and (Q), as well as the exact value from TBA.}
\label{tab:wynn0}
\end{table}

\begin{figure}[th]
\subfigure[The function $g_0$]{\scalebox{0.7}{\includegraphics{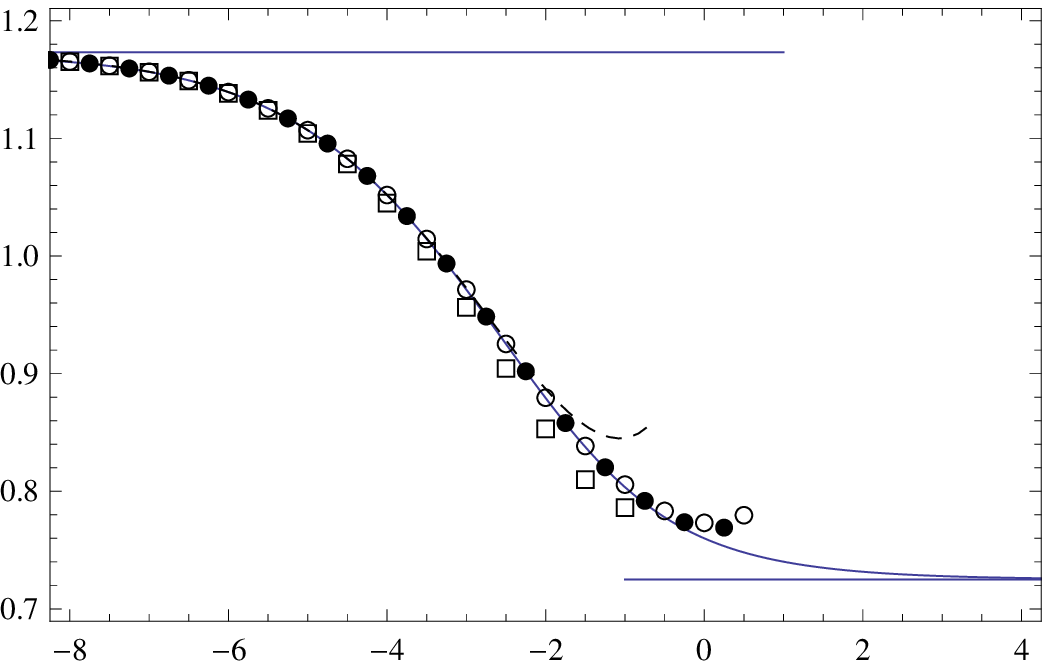}}
\label{fig:yl1}}
\hfill
\subfigure[The functions $g_1$]{\scalebox{0.7}{\includegraphics{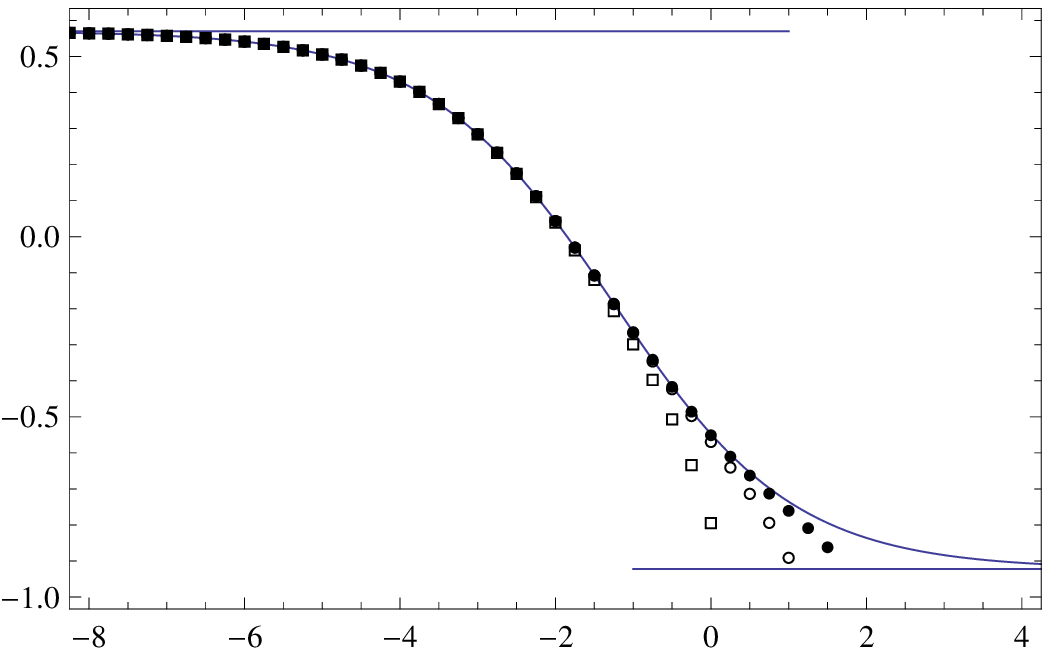}}
\label{fig:yl2}}
\hfill
\caption{The TCSA estimates for the 
$g$-functions $g_0$ and $g_1$ level 0
  (open squares),
20 (open circles), and extrapolated (solid circles), compared with the
exact results from TBA (solid lines). 
Also shown are the UV and IR values as straight lines and the level 18
result for $g_0$ from the method of \cite{Dorey:2000} (dashed line).
}
\label{fig:yl}
\end{figure}

Although the agreement with the functions $Y_0$ and $Y_1$ as plotted
in figure 1.\ of \cite{BLZ:9607099} is impressive, this is in main due
to the dominance of the functions $Y_i$ by the boundary free energy
term which means that the $Y$ functions grow rapidly and are hard to show
on this particular plot.
To remove this factor, 
we instead plot the 
the $g$-functions $g_0(h R^y)$ and 
$g_1(h R^y)$
which interpolate the UV and IR values of $g_\alpha$ and
$Z_\alpha$ respectively, using \cite{Dorey:2000}
\be 
2\pi f_B R = - \pi | 2 h/h_c|^{5/6}
\;.
\ee
In figure \ref{fig:yl} we show tha 
TCSA estimates (for truncation levels 0 and 20) and TBA values of
$g_0$ and $g_1$, along with the result of a Pad\'e approximant extrapolation in
truncation level to $N=\infty$.
\begin{figure}[htb]
\centering{\scalebox{0.7}{\includegraphics{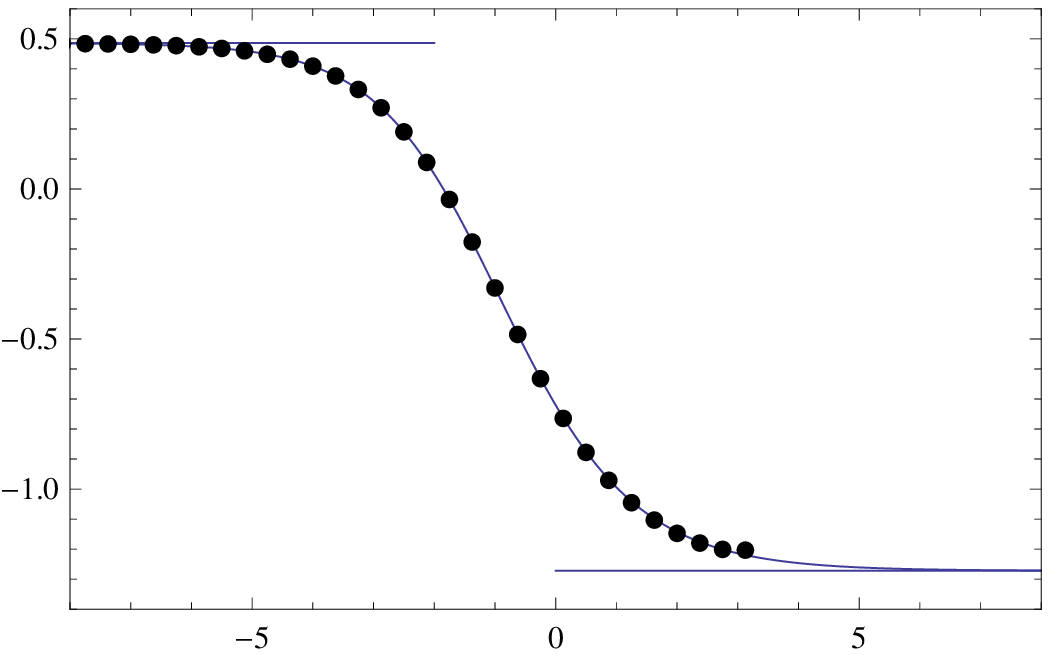}}}
\caption{The TCSA estimate of the ratio $g_1/g_0$ from truncation at
  level 20, together with the
exact results from TBA (solid lines). 
Also shown are the UV and IR values as straight lines.
}
\label{fig:g01}
\end{figure}
As can be seen in figure \ref{fig:yl1}, the estimate of $g_0$ and $g_1$ from
the level zero truncation is already quite good and convergence
improves rapidly with increasing truncation level, so that for level
20 it is qualitatively accurate up to $\log h \simeq -0.5$.
The main error seems to come from errors in the extensive boundary
energy per unit length which can be attributed to an effective
coupling constant renormalisation in the TCSA method
\cite{GW:BTCSARG}. Rather than attempt to incorporate these effects
in the differential equation, we can remove the error from the
boundary free energy by looking at the ratio of the $g$-functions,
which is plotted in figure \ref{fig:g01}.

We can attempt to improve the accuracy of our estimates by using an
extrapolation method to estimate the 
$N\to\infty$ limit of the TCSA data. We have used two ways of approaching
the $N\to\infty$ (described in section \ref{sec:YL2})
limit and three different extrapolation methods.
In all cases, 
all three extrapolation methods gave the similar results, with Wynn's method
\cite{Wynn} based on Pad\'e approximants being the best. 
The two different ways of approaching the limit, while very similar
for $g_0$ and $g_1$, differed considerably for some higher excited
states.
The simplest method, considering the level $N$
approximation to be given by the full TCSA expression \eref{eq:1},
failed to be effective for fields of weight greater than 2,
and we had to use a second approximation to get an effective
extrapolation to $N=\infty$.
We illustrate this in table \ref{tab:wynn0}  with the TCSA estimates
of $g_0$ and $g_1$ at $\log(h)= -2$ for $0\leq N\leq 20$ together with
the two Pad\'e extrapolations labelled $\infty^{(P)}$ and
$\infty^{(Q)}$  and the TBA value. We also include the values of
$g_2$, $g_3$, $g_4$ and $g_5$ which will be discussed in the next
section, together with a discussion of the two extrapolations.

To conclude this section, we discuss the comparison with the previous
TCSA estimation of $g_0$ in \cite{Dorey:2000}. In that paper, $g_0$
was estimated from the scaling with the strip width of the cylinder
partition function calculated as a trace over the space of states on
the strip. This was 
a rather convoluted calculation which nevertheless showed good
agreement with the exact functions. In figure \ref{fig:yl1} we include
the old TCSA estimate from the level 18 truncation 
(using the exact boundary free energy per unit length as
shown in \cite{Dorey:2000}); as can be seen, the new
method is much better. The old TCSA estimate could be improved by
fitting the boundary free energy term to account for the renormalisation of
the coupling constant, as can the new TCSA estimate by extrapolation
in $N$, and the new method again gives better results.

\sect{Higher excited states in the Lee-Yang model}
\label{sec:YL2}

We can continue to apply the same method to the higher excited states
in the Lee-Yang model. The next four states are given in table
\ref{tab:lystates} and three of the corresponding fields are given by simple derivatives
of the primary field $\varphi(z,\bar z)$.  We can calculate these by
directly taking the derivative of the one-point function on the
perturbed disc of $\varphi(z,\bar z)$, which we discuss in appendix \ref{sec:dp}.
The fourth excited state corresponds to the field
$T(z)\bar T(\bar z)$ and we discuss its matrix elements in appendix \ref{sec:ttb}.
The results for the excited state functions $g_2$ to $g_5$ at 
$\log h = -2$  in table \ref{tab:wynn0}.  As can be seen, these  
are convergent for the first three states but oscillatory and
divergent for $g_3$ to $g_5$. The behaviour of these divergent cases
is of an increasing oscillating series, typical of a divergent
perturbation expansion. 

This immediately suggests that one might be
able to re-sum the divergent series and obtain useful results. 
We have
tried Euler, Cesaro and Pad\'e extrapolation (using Wynn's method
\cite{Wynn}) and these all gave similar results. They all improve the
results ofr $g_0$ to $g_2$, are all able to
produce a result correct to a few percent for the divergent series
$g_3$ and $g_4$ but all fail on $g_5$. As a consequence we considered
different ways to approach $N\to\infty$ limit, that is we constructed
a different TCSA approximation at each level which would agree for
convergent series, but could give better results for divergent
series. Looking at the matrix elements $\cev \alpha \cO_i \vec \beta$
of the fields $\cO_i$ it seemed 
that these were organised by the total excitation level of the states
$\vec\alpha$ and $\vec\beta$ rather than the maximum level.
We consequently defined two series as follows.
We first denote the contributions at excitation level $n$ to the level
$N$ TCSA approximation $\vec{\Psi_N}$ to the state $\vec{\Psi(1)}$ by
$\vec{\psi_n}$, 
so that
\be
\vec{\Psi_N} = \sum_{n=0}^N \vec{\psi_n}
\;,
\ee
For any operator $\cO$, we then define two series of polynomials
\bea
    P^{\cO}_N(\lambda)
&=& \sum_{m=0,n=0}^N \lambda^{max(m,n)}\,\cev{\psi_m}\,\cO\,\vec{\psi_m}
\;,
\nonumber\\
    Q^{\cO}_N(\lambda)
&=& \sum_{m+n=0}^N \lambda^{m+n}\,\cev{\psi_m}\,\cO\,\vec{\psi_m}
\;.
\eea
The direct TCSA approximation at level $N$ is given by $P^{\cO}_N(1)$. 
If the TCSA approximation converges in $N$ then both
$P^{\cO}_N(1)$ and $Q^{\cO}_N(1)$ converge to the same answer;
if the TCSA expression diverges then the two polynomials can differ
considerably and we can take the Pad\'e
approximants to $P_N(\lambda)$ and $Q_N(\lambda)$ at $\lambda=1$ and
see if they provide useful estimates of the answer.
As an example, we consider $\cO = (-11/5)T(0)\bar T(0)$ and the truncation to
level 12 at $h=1$. The polynomials are
\bea
 P^{T\bar T}_{12} 
&\!=\!&
  657.76 - 1077.2 \,\lambda^2 + 3234.6 \,\lambda^4 - 5513.1 \,\lambda^6 + 
 8256.0 \,\lambda^8 - 11183 \,\lambda^{10} + 14552 \,\lambda^{12}
\;,\nonumber\\
 Q^{T\bar T}_{12}
&\!=\!&
  657.76 - 3671.3 \,\lambda^2 + 8657.4 \,\lambda^4 - 14697 \,\lambda^6 + 
 21200 \,\lambda^8 - 27983 \,\lambda^{10} + 35138 \,\lambda^{12}
\;,\nonumber
\eea
\be
  P^{T\bar T}_{12}(1) =   8926.8
\;,\;\;
  Q^{T\bar T}_{12}(1) =   19301.
\;.
\ee
We see that neither of these is close to the exact result of -113.6. 
Instead we take 
the Pad\'e approximants
\bea
   p^{T\bar T}_{12}(\lambda)
&=&
\frac
 {657.76  - 234.64 \,\lambda^2 + 1388.4 \,\lambda^4  -  1277.3
   \,\lambda^6}
 {1 +  1.2810 \,\lambda^2  - 0.70901 \,\lambda^4  -  1.0206
   \,\lambda^6}
\;,
\,\nonumber\\
   q^{T\bar T}_{12}(\lambda)
&=& 
\frac
     {657.76  - 2011.4\,\lambda^2 +  766.95 \,\lambda^4  -  
     127.37 \,\lambda^6}
     {1 + 2.5237 \,\lambda^2 + 2.0903 \,\lambda^4  + 0.60087 \,\lambda^6}
\;,\nonumber\eea
and find the Pad\'e approximant estimates and the exact (TBA) answer are
\be
   p^{T\bar T}_{12}(1) = 968.9
\;,\;\;
   q^{T\bar T}_{12}(1) = -114.9
\;,\;\;
  \cG_4^{\mathrm{TBA}} = -113.6
\;.
\ee
We see that the $Q$-approximation is reliable and the
$P$-approximation is not.
This is repeated for the other divergent cases of $g_3$ and $g_4$ -
the $Q$-estimate is always better than the $P$-estimate.
It is worth noting that for the convergent cases there is very little
difference between the two, but that the $P$-estimate is very slightly
better than the $Q$-estimate since the $P$-polynomial includes more (convergent)
matrix elements than the $Q$-polynomial.

\begin{figure}[thb]
\centering{{$  
 \begin{picture}(300,195)
  \put(0,0){{\scalebox{0.9}{\includegraphics{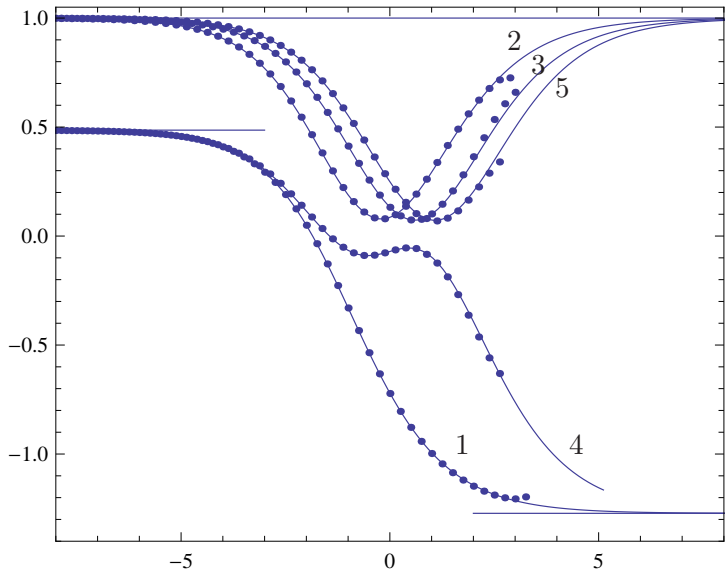}}}}
  \put(0,0){
     \setlength{\unitlength}{0.9pt}\put(0,0){
     \put(210,220)   {$2$}
     \put(220,210)   {$3$}
     \put(230,200)   {$5$}
     \put(188,50)   {$1$}
     \put(236,50)   {$4$}
     }\setlength{\unitlength}{1pt}
            }
  \end{picture}
$}}
\label{fig:zz}
\caption{The extrapolations of the TCSA estimates for $g_i/g_0$
  plotted against $\log(h)$ for
  $i=1,2,3,4$ and 5 (labelled by $i$). Also shown are the TBA values (solid lines) and
  the only  allowed values in the IR and UV (straight lines).
} 
\label{fig:zzz}
\end{figure}

We show the resulting plots for the ratios $g_i/g_0$ of the $Q$-extrapolated
TCSA results at level 20 
in figure \ref{fig:zzz}. We see that the ratios,
which are the physical content of the $g$-functions, 
are all well described by the $Q$-extrapolated TCSA data up to
approximately the same value of $\log h\simeq 2.5$, which is the value at which the
TCSA Hamiltonian on a strip also fails to give an accurate description
for the 5th excited state of the spectrum.

\section{Conclusions}

We have found a new and effective way to calculate the overlaps of
perturbed boundary states with bulk states giving excited state
$g$-functions, or equivalently $T$-functions. We have studied these in
some detail in the Lee-Yang model (in which the $T$-functions are
identically equal to the $Y$-functions) and shown that the estimates
are good for the whole range of values for which the TCSA method on
the strip is applicable. 
Even using a truncation to just two states, the results are better
than second order perturbation theory.
It is disappointing, although perhaps not surprising, that the TCSA
estimates computed directly are divergent for the states of conformal
weight greater than 2, but we have found an efficient extrapolation
method which works effectively for all the states we have considered.

This new TCSA method can be applied to any boundary perturbation for which the
boundary couplings are known and is not restricted to integrable
models, as was the case here. 
The method also has immediate scope for generalisation to include massive,
or bulk, perturbations, which should help check any conjectured TBA
equations for the bulk excited state $g$-functions which are
complicated by the extra infinite series required to account for the
bulk corrections \cite{DT,P}.
It can also be applied to study defect matrix elements (which
are boundary state overlaps in the double model through the usual
mirror trick) and calculate the reflection and transmission
coefficients defined in \cite{quella} 
and help clarify the nature of the endpoints of the defect flows studied
in \cite{marton}.

\section*{Acknowledgements}


GMTW would like to thank ELTE for hospitality at the start of this
project and STFC grant
ST/G000395/1 for support and Zolt\'an Bajnok, Patrick Dorey, Roberto Tateo and Ingo
Runkel for useful discussions.
GT was partially supported by the Hungarian OTKA grant K75172.
The TCSA calculations were performed using Mathematica \cite{mathematica}.

\label{sec:acc}

\makeatletter

\providecommand{\tabularnewline}{\\}

\makeatother

\appendix

\newpage
\section{The Lee-Yang conformal data}
\label{app:ly}

The coefficients appearing in the boundary states \eref{eq:bss} are
\bea
&&  Z_\Phi 
= \left| \frac{\sqrt 5 - 2}{\sqrt 5}\right|^{1/4}
= 0.570017...
\;,\;\;
  Z_\Bone 
= -\left| \frac{\sqrt 5 +1}{2\sqrt 5}\right|^{1/4}
= -0.922307...
\;,
\\
&&
  g_\Phi^{\vphantom\phi}
= \left| \frac{\sqrt 5 + 2}{\sqrt 5}\right|^{1/4}
= 1.17319...
\;,\;\;\;\;
  g_\Bone^{\vphantom\phi}
= \left| \frac{\sqrt 5 -1}{2\sqrt 5}\right|^{1/4}
= 0.725073...
\;.
\eea
The various structure constants used in this paper are
\bea
\begin{array}{rclcl}
&&c_{\phi\phi}{}^1
= -1
\;,
&&\ds
  c_{\phi\phi}{}^\phi
= - \left|\frac{1+\sqrt 5}{2}\right|^{1/2}\!\!\!\!\cdot\alpha
= -1.98338..
\;,
\\[5mm]
&&\ds
  {}^{(\Phi)}\!B_\Phi^\phi
= \left|\frac{5 + \sqrt 5}{2}\right|^{1/2}\!\!\!\!\cdot\alpha
= 2.96585..
\;,
&&\ds
  {}^{(\Phi)}\!B_\Phi^1
= \left|\frac{1 + \sqrt 5}{2}\right|^{3/2}
= 2.05817..
\;,
\end{array}
\eea
where
\be
  \alpha 
= \left| \frac{\Gamma(1/5)\Gamma(6/5)}{\Gamma(3/5)\Gamma(4/5)}
  \right|^{1/2}
= 1.55924..
\;.
\ee
Finally, the one-point functions of $\varphi(w,w)$ on the UHP are
given in terms of $r=|w|$ and $\theta=\mathrm{arg}(w)$ as
\be
\begin{array}{rcl}
\cav\,\varphi(w,\bar w)\,\vac
&=&
- \,{}^{(\Phi)}\!B_\varphi^1\;
  (2r\sin\theta)^{2/5}
\\
\cav\,\varphi(w,\bar w)\,\vec{\,\phi\,}
&=&
- \,{}^{(\Phi)}\!B_\varphi^\phi\;
  r^{2/5}\,(2r\sin\theta)^{1/5}
\\
\cev{\,\phi\,}\,\varphi(w,\bar w)\,\vac
&=&
- \,{}^{(\Phi)}\!B_\varphi^\phi\;
  (2r\sin\theta)^{1/5}
\\
\cev{\,\phi\,}\,\varphi(w,\bar w)\,\vec{\,\phi\,}
&=&
- c r^{2/5} (2\sin\theta)^{1/5} F(\cos^2(\theta)) 
\;,\;\;
%
%
\end{array}
\label{eq:1ptfns}
\ee
where $c=
\tfrac{\sqrt{1 {+} \sqrt 5} }{\textstyle 2^{1/10}}
\frac{    \Gamma(\tfrac 15) \Gamma(\tfrac{11}{10}) }
     { \sqrt\pi \Gamma(\tfrac 45)}$
and 
$F(x) = 
 F(\tfrac{1}{10}, \tfrac{3}{10}, \tfrac{1}{2}, x)$.

\section{Perturbation theory}
\label{sec:pt}

It is straightforward to obtain one or two orders of the perturbation
epansions of the functions $\cG_i$. For the $g$-function $\cG_0$ this
was given in \cite{BLZ:9607099} and the conformal theory perturbation 
discussed in \cite{Dorey:2000}. Here we write them in
terms of $h R^{6/5}$ rather than $h (2\pi R)^{6/5}$:
\be
\cG_0 = g_\Phi \Big( 1 + 
\alpha_{1} h R^{6/5} + \alpha_2 (h R^{6/5})^2 + \ldots
\;\Big)\;\;,
\nonumber\ee\be
\alpha_1 = (2\pi)\, 
\left| \frac{\sqrt 5 \,\Gamma(2/5)\Gamma(6/5) }{2\cos(\pi/5)\Gamma(4/5)^2}
\right|^{1/2}
\;,\;\;
\alpha_2 = \alpha_1^2 / \sqrt 5
\;.
\ee
The function $\cG_1$, the disk partition function, can be expanded out
to third order as
\be
\cG_1
 = Z_\Phi
  \Big(
  1 
\;-\; \tfrac 12(hR^y)^2 (2\pi)^2
      \frac{2^{2/5}\,\Gamma(7/10)}{\sqrt\pi\,\Gamma(6/5)}
\;+\;\tfrac 16 (hR^y)^3 (2\pi)^3 \frac{\Gamma(13/10)}{\Gamma(11/10)^3}\,c_{\phi\phi}{}^\phi
\;+\;\ldots\;
  \Big)
\ee
Similarly, $\cG_2,\cG_3$ and $\cG_5$ can be expanded to second order
and $\cG_4$ to third order based on conformal perturbation theory;
beyond these orders the integrals become difficult to evaluate
exactly, although use of the Y-function relation
\eref{eq:yeqn} does allow one to
calculate one further order for $\cG_0$ and three further orders for
$\cG_1$ as the coefficients in the perturbation expansion are severely
constrained by this relation \cite{BLZ:9607099}, and so obtain
\bea
\cG_0 &=& g_\Phi\Big( 
1 + \alpha_1 x + \frac{1}{\sqrt 5} (\alpha_1)^2 x^2
+ \frac{\sqrt 5 - 1}{10} (\alpha_1)^3 x^3 + \ldots\;\Big)
\\
\cG_1 &=& Z_\Phi\Big(
1 + \beta_2 x^2  + \beta_3 x^3 
-\frac{(\beta_2)^2}{\sqrt 5} x^4
+ \frac{5 - 3\sqrt 5}{10} \beta_2\beta_3 x^5
\nonumber\\
&&\qquad\qquad\qquad\qquad\qquad\qquad
- \frac{ (1 + \sqrt 5)(\beta_2)^3 + 2\sqrt 5(\beta_3)^2}{10} x^6 +
\ldots\;\Big)
\eea
where $x = h R^{6/5}$. These are also shown in figure \ref{fig:level0}

It is rather harder to calculate the $z$-dependent function
$\vev{\varphi(z,\bar z)}$ as the integrals again become harder. To
first order we have (for a disk of radius 1)
\be
\vev{\varphi(z,\bar z)}
= Z_\Phi\,
 (1 {-} |z|^2)^{2/5}\,
 \left(
   {}^{(\Phi)}\!B_\varphi^1 
 \,+\, (2\pi) h 
   {}^{(\Phi)}\!B_\varphi^\phi\,(1{-}|z|^2)^{-1/5}\,
   {}_2F_1(-\tfrac 15,-\tfrac 15;1;|z|^2)
 \,+\,\ldots\,
 \right)
\;.
\ee

\section{$\varphi(z,\bar z)$ expectation values}
\label{sec:dp}

Of the first four excited states given in table \eref{tab:lystates},
three are derivatives of the primary field $\varphi(z,\bar z)$.  We
can calculate the overlaps with these states by
directly taking the derivative of the one-point function of
$\varphi(z,\bar z)$. 
Here we show how this is defined in our TCSA approach.
We use the coordinates $z$ on the unit disk and $w = r\exp(i\theta)$
on the UHP, related by \eref{eq:iota}.
In these coordinates 
\be
z = \frac{1 - r^2 + 2 i r \cos\theta}{1 + r^2 + 2 r \sin\theta}
\;,\;\;
|z|^2
= \frac{1 + r^2 - 2  r \sin\theta}{1 + r^2 + 2 r \sin\theta}
\;,\;\;
\left|\frac{\rd w}{\rd z}\right|
= \tfrac1{2}(1 + r^2 + 2 r \sin\theta)
\;.
\ee 
The general form of the expectation value of an operator $\cO$ on the
disk is given by \eref{eq:expect} and the transformation property of the
primary field $\varphi(z,\bar z)$ is given by \eref{eq:phimap}, so
that
\bea
\vev{\;\varphi(z,\bar z)\;}_{\mathrm{disk}}
&=& 
\left(\tfrac 12(1 + r^2 + 2 r \sin\theta)\right)^{-2/5}
\vev{\;\varphi(w,\bar w)\;}_{\mathrm{UHP}}
\nonumber\\
&=&
- Z_{\Phi}
\,
\left(\tfrac 12(1 + r^2 + 2 r \sin\theta)\right)^{-2/5}
\cev{\Psi(1)}\varphi(re^{i\theta},re^{-i\theta})\vec{\Psi(1)}
\;.
\eea
We then used 
\be
\varphi(re^{i\theta},re^{-i\theta})
= r^{2/5} r^{L_0+\bar L_0}
\varphi(e^{i\theta},e^{-i\theta})
r^{-L_0-\bar L_0}
\;,
\ee
to arrive at
\be
\vev{\;\varphi(z,\bar z)\;}_{\mathrm{disk}}
= 
- Z_{\Phi}
\,
\left(\tfrac 12(r + 1/r + 2 \sin\theta)\right)^{-2/5}
\cev{\Psi(1)}r^{L_0+\bar
  L_0}\varphi(e^{i\theta},e^{-i\theta})\,r^{-L_0-\bar L_0}\vec{\Psi(1)}
\;.
\ee
The matrix elements of $\varphi(e^{i\theta},e^{-i\theta})$ can be
calculated in the usual way from the  highest weight
matrix elements \eref{eq:1ptfns}. Given these, the TCSA estimate takes the form
\blank{
, the
one-point functions of $\varphi(w,\bar w)$ on the UHP are given in
\eref{eq:1ptfns} and the general form of the state $\vec{\Psi(1)}$ is
given by \eref{eq:Psiform}.
Putting these together we arrive at the TCSA form of the expectation
value of $\varphi(z,\bar z)$, expressed in terms of $r$ and $\theta$:
\be
\vev{\,\varphi(z,\bar z)}
=
\;.
\label{eq:expect2}
\ee
The problem is that the TCSA expression for $\vev{\varphi(z,\bar z)}$
takes a form akin to a Fourier series and the derivatives of
$\vev{\varphi(z,\bar z)}$ are not necessarily continuous.  It is well
known that if a function has discontinuous first derivatives then the
second and higher derivatives of its Fourier series will not converge.
While we do not exactly have the situation of a Fourier series
expansion, the result is very similar.
and the form of 
the TCSA approximation to
the perturbed one-point
function at at truncation level $N$ is
}
\be
  \vev{ \varphi(z,\bar z) }_{\Phi(h)}^{(N)}
= \left( 1 - |z|^2 \right)^{2/5}
  \left[ 
  \alpha^N_{0} 
+ \alpha^N_{1} (\sin\theta)^{-1/5}
+ \alpha^N_{2} f_{2}(\theta) 
+ \beta^N_{2} f_{3}(\theta)
  \right]
\;,
\ee
where 
\be
f_{2}(\theta) =  \sin(\theta)^{-1/5}\,
                  F(1/10, 3/10; 1/2; \cos^2\theta)
\;,\;\;
f_{3}(\theta) =
\sin(\theta)^{-1/5}\frac{\rd}{\rd\theta}[(\sin\theta)^{2/5}f_{2}]
\;,
\ee
and the coefficient functions have the form
\bea
\alpha^N_{i} = \sum_{j=0}^{2N} \alpha^{N,j}_i(r) \cos(2j\theta)
\;,\;\;
\beta^N_{i} = \sum_{j=0}^{2N} \beta^{N,j}_i(r) \sin(2j\theta)
\eea
For the level 0, two-state truncation with $\vec{\Psi(1)} = a\vac +
  b\vec\phi$, these functions are
\be
\alpha_0^0 = {}^{(\Phi)}\!B_\varphi^{\one}\,Z_\Phi\,\cdot\,a^2
\;,\;\;\;\;
\alpha_0^1 = {}^{(\Phi)}\!B_\varphi^{\phi}\,Z_\Phi\,2^{4/5}\,
 (r^{1/5}+r^{-1/5})\,\cdot
 \,a\,b
\;,\;\;
\nonumber\ee\be
\alpha_2^0 = 2^{-1/5}\,c\,Z_\Phi\,\cdot\,b^2
\;,\;\;\;\;
\beta_2^0 = 0
\;.
\ee
The map $z\to w$ is conformal and the Laplacian in these coordinates is 
\be
\partial\bar\partial =
\frac{\left({ r^2 + 2 r \sin\theta + 1}\right)^2}{16}
\left(
\frac{\partial^2}{\partial r^2} + \frac{1}{r}\frac{\partial}{\partial r}
+ \frac{1}{r^2}\frac{\partial^2}{\partial\theta^2}
\right)
\;.
\ee
While the functions $\alpha^N_x$ converge in $N$, the repeated action 
of $\partial\bar\partial$ leads to a failure of convergence, which in
this case is happens first for $g_3$ with two applications of
$\partial\bar\partial$.

\section{ $T\bar T$ expectation values}
\label{sec:ttb}

The field $T\bar T$ is quasi-primary and transforms under the Mobius
map \eref{eq:iota} as
\be
  T(z)\bar T(\bar z) \mapsto \left| w'(z) \right|^4\, T(w)\bar T(\bar
  w)
\;,
\label{eq:Tiota}
\ee
which gives the expectation value on the perturbed disk as
\bea
\vev{ T(z)\bar T(\bar z) }_{disk}
&=& \left( \frac{1 + r^2 + 2 r \sin(\theta) }{2} \right)^4
\vev{ T(re^{i\theta}) \bar T(re^{-i\theta}) }_{UHP}
\nonumber\\
&=&
- Z_\Phi
  \,
  \left( \frac{1 + r^2 + 2 r \sin(\theta) }{2} \right)^4
  \,\cev{\Psi(1)}  T(re^{i\theta}) \bar T(re^{-i\theta}) \vec{\Psi(1)}
\eea
The product $T(r e^{i\theta}) T(r e^{-i\theta})$ is regular and its
matrix elements can be worked out in the truncated conformal space
using the normal ordering relation
identity appropriate for the TCSA method,
\bea
 T(z) T(w)
&=& T_{\leq 0}(z) T(w) + T(w) T_{>0}(z)
\nonumber\\
&&\;\;+ \frac{c/2}{(z-w)^4} 
+ \left[\frac{2}{(z-w)^2} - \frac{2}{z^2}\right]T(w)
+ \frac{w}{z^2(z-w)}T'(w)
\;,
\label{eq:TTope}
\eea
where 
$T_{\leq 0}(z) = \sum_{m\leq 0} L_m z^{-m-2}$ 
and
$T_{> 0}(z) = \sum_{m> 0} L_m z^{-m-2}$.
This choice of ordering means that the matrix elements of $T(e^{i\theta})T(e^{-i\theta})$ in the
truncated space can be
calculated exactly using the expansion of the  right-hand-side of
\eref{eq:TTope} in the truncated conformal space.
To work out the level zero TCSA approximation, we just need the
expectation value in a highest weight state,
\bea
\cev h T(z)T(w) \vec h
&=& \frac{c/2}{(z-w)^4} + \frac{2h}{zw(z-w)^2} + \frac{h^2}{z^2w^2}
\;,
\\
\cev h T(re^{i\theta})T(r^{-i\theta}) \vec h
&=& \frac{c/2}{(2r\sin\theta)^2}
    \left[
     (1 + \tfrac{16 h(1{-}2h)}{c})
 \;+\;
    \cos^2\theta \tfrac{16h(1{-}4h)}{c}
 \;+\;
    \cos^4(\theta) \tfrac{32}{c}
    \right]
\;.
\nonumber
\eea
Putting this all together, we find the level zero TCSA approximation
to the disk one-point function of the normalised field is
\be
  \frac{1}{Z_\Phi}\,
  \vev{ \;\tfrac{2}{c}\,T(z)\bar T(\bar z)\; }_{disk}^{level\,0}
= (1 - |z|)^{-4}
  \left[ 
  a^2 + b^2\,\frac{1}{55}(1  - 72\cos^2\theta + 16\cos^4\theta)
  \right]
\;.
\ee
As is immediately obvious, this does not have the expected
rotationally-invariant form $f(|z|)$, since the TCSA method breaks rotational
invariance.

\blank{
We also give the level-two TCSA approximation. We shall take
$\vec{\Psi(1)}$ to be the state
\be
\vec{\Psi(1)} = a\vec\phi + b\vac + c L_{-1}\vec\phi +
 d L_{-2}\vec\phi + e  L_{-2}\vac 
\;.
\ee
Since expectation in the state $\vec{\Psi(1)}$ diverges as the
truncation level is raised, it is useful to put in a parameter
$\lambda$ which measures the excitation level of the states 
contributing to $\vec{\Psi(1)}$; this will enable us to use Wynn's
method and use a Pad\'e approximant to estimate the $N\to\infty$ limit
of the TCSA expectation value. In this case, we give the level 2
expectation value in the state 
\be
\vec{\Psi(1)} = \left[a\vec\phi + b\vac\right] +
\lambda\left[ c L_{-1}\vec\phi \right]+
\lambda^2\left[ d L_{-2}\vec\phi + e  L_{-2}\vac \right]
\;,
\ee
which is
\bea
&&  \frac{1}{Z_\Phi}\,
  \vev{ \;\tfrac{2}{c}\,T(z)\bar T(\bar z)\; }_{disk}^{level\,2}
\nonumber\\
&=& (1 - |z|)^{-4}\Big(
  \left[ 
  a^2 + b^2\,\frac{1}{55}(1  - 72\cos^2\theta + 16\cos^4\theta)
  \right]
\nonumber\\
&+&\lambda^2\left[ \right]
\nonumber\\
&+&\lambda^4\left[ \right]
\,\Big)
\;.
\eea
Taking the Pad\'e approximant for $\lambda\to 1$, this gives
\bea
&&  \frac{1}{Z_\Phi}\,
  \vev{ \;\tfrac{2}{c}\,T(z)\bar T(\bar z)\; }_{disk}^{level\,2}
\nonumber\\
&\simeq&
\eea 
}

\section{TBA equations in the Lee-Yang model}

The scaling Lee-Yang model is the perturbation of the conformal minimal
model $\mathcal{M}_{2,5}$ by the relevant field $\Phi_{1,2}$:
\[
\mathcal{A}_{\mathrm{SLYM}}=\mathcal{A}_{2,5}+ih\int dzd\bar{z}\Phi_{1,2}(z,\bar{z})
\]
It has a single particle of mass $m$ in the spectrum with the two-particle
$S$ matrix 
\[
S=\frac{\sinh\theta+i\sin\frac{2\pi}{3}}{\sinh\theta-i\sin\frac{2\pi}{3}}
\]
The ground state TBA of the SLYM has the form \cite{Zamolodchikov:1989cf}
\begin{eqnarray*}
\epsilon(\theta) & = & mL\cosh(\theta)-\int_{-\infty}^{\infty}\frac{d\theta'}{2\pi}\varphi(\theta-\theta')\log(1+\mathrm{e}^{-\epsilon(\theta')})\\
E_{0}^{\mathrm{TBA}}(L) & = & -\int_{-\infty}^{\infty}\frac{d\theta}{2\pi}m\cosh\theta\log(1+\mathrm{e}^{-\epsilon(\theta)})
\end{eqnarray*}
where 
\[
\varphi(\theta)=-i\frac{d}{d\theta}\log S(\theta)=-\frac{\sqrt{3}\cosh\theta}{\cosh^{2}\theta-1/4}
\]
The excited state levels can be determined by solving the equations
\cite{Dorey:1996re,BLZ:9607099} 
\begin{eqnarray}
\epsilon(\theta) & = & mL\cosh(\theta)+\sum_{j=1}^{k}\log\frac{S(\theta-\theta_{j})}{S(\theta-\theta_{j}^{*})}\label{eq:lyexcitedtba}\\
 &  & -\int_{-\infty}^{\infty}\frac{d\theta'}{2\pi}\varphi(\theta-\theta')\log(1+\mathrm{e}^{-\epsilon(\theta')})\nonumber \\
\mathrm{e}^{-\epsilon(\theta_{j})} & = & -1\nonumber \\
E^{\mathrm{TBA}}(L) & = & -i\sum_{j=1}^{k}m\left(\sinh\theta_{j}-\sinh\theta_{j}^{*}\right)-\int_{-\infty}^{\infty}\frac{d\theta}{2\pi}m\cosh\theta\log(1+\mathrm{e}^{-\epsilon(\theta)})\nonumber 
\end{eqnarray}
where $k$ is the number of particles present in the state with the
singularity positions $\theta_{j}$ corresponding to their rapidities.
For large volumes, the integral terms can be dropped and a straightforward
calculation shows that the singularity positions have the form
\[
\theta_{j}=\tilde{\theta}_{j}+i\frac{\pi}{6}+O(\mathrm{e}^{-mL})
\]
where the particle rapidities $\tilde{\theta}_{j}$ solve the Bethe-Yang
equations
\begin{eqnarray*}
mL\sinh\tilde{\theta}_{j}+\sum_{l\neq j}\delta\left(\tilde{\theta}_{j}-\tilde{\theta}_{k}\right) & = & 2\pi I_{j}\\
\mbox{where }\quad S(\theta) & = & -\mathrm{e}^{i\delta(\theta)}
\end{eqnarray*}
with $I_{j}\in\mathbb{Z}$ (for $k$ odd) or $I_{j}\in\mathbb{Z}+\frac{1}{2}$
(for $k$ even) labelling the particular state in question.

The $Y$ function of the given level can be defined as 
\[
Y(\theta)=\mathrm{e}^{\epsilon(\theta)}
\]
and it satisfies the functional relations \cite{Zamolodchikov:1991et}
\begin{equation}
Y\left(\theta+i\frac{\pi}{3}\right)Y\left(\theta-i\frac{\pi}{3}\right)=1+Y\left(\theta\right)\label{eq:yeqn}
\end{equation}

These equations can be solved by iteration: the Bethe-Yang equations
give a starting point for the position of the singularities, which
can then be used to iterate the system (\ref{eq:lyexcitedtba}). For
a given position of singularities the function $\epsilon$ can be
updated by iterating the integral equations, and then the positions
of the singularities can be refined by solving the equations 
\[
\Re e\,(\mathrm{e}^{-\epsilon(\theta_{j})}+1)=0\qquad\Im m\,(\mathrm{e}^{-\epsilon(\theta_{j})}+1)=0
\]
for the $2k$ real variables given by the real and imaginary parts
of $\theta_{j}$. This can be accomplished by the multidimensional
Newton method; the $2k\times2k$ derivative matrix of the equations
with respect to the variables can be expressed by taking the derivative
of the integral equation in (\ref{eq:lyexcitedtba}).

To describe the states in the minimal model, we consider the so-called
kink limit of the above TBA system describes a chiral (right moving)
version of the $c=-22/5$ Lee-Yang minimal model. The variable $\theta$
must be redefined by the shift
\[
\theta\rightarrow\theta-\log\frac{2}{mL}
\]
Taking the $mL\rightarrow0$ limit one arrives at the TBA system
\begin{eqnarray}
\epsilon(\theta) & = & \mbox{e}^{\theta}+\sum_{j=1}^{k}\log\frac{S(\theta-\bar{\theta}_{j})}{S(\theta-\bar{\theta}_{j}^{*})}-\int_{-\infty}^{\infty}\frac{d\theta'}{2\pi}\varphi(\theta-\theta')\log(1+\mathrm{e}^{-\epsilon(\theta')})\nonumber \\
\mathrm{e}^{-\epsilon(\bar{\theta}_{j})} & = & -1\nonumber \\
E(L) & = & -\frac{\pi c_{\mathrm{eff}}}{6L}\label{eq:massless_lytba}
\end{eqnarray}
where the only remaining $\theta_{j}$ are those for which the limit
\[
\bar{\theta}_{j}=\lim_{L\rightarrow\infty}\theta_{j}-\log\frac{2}{mL}
\]
is finite, and the effective central charge of the state can be expressed
as 
\[
c_{\mathrm{eff}}=\frac{12i}{\pi}\sum_{j=1}^{k}\left(\mathrm{e}^{\bar{\theta}_{j}}-\mathrm{e}^{\bar{\theta}_{j}^{*}}\right)+\frac{6}{\pi^{2}}\int_{-\infty}^{\infty}d\theta\mathrm{e}^{\theta}\log(1+\mathrm{e}^{-\epsilon(\theta)})=c-24\Delta
\]
where $\Delta$ is the dimension of the operator creating the state
and the central charge is $c=-22/5$. 

The massless ground state $Y$-function
\[
Y_{0}(\theta)=\epsilon_{0}(\theta)
\]
 can be found simply by iterating 
\[
\epsilon_{0}(\theta)=\mbox{e}^{\theta}-\int_{-\infty}^{\infty}\frac{d\theta'}{2\pi}\varphi(\theta-\theta')\log(1+\mathrm{e}^{-\epsilon_{0}(\theta')})
\]
which converges very fast. The ground state is created by the operator
$\varphi$ with weight $-1/5$, with $c_{\mathrm{eff}}=2/5$.

For a reliable numerical computation of the massless excited state
$Y$-function we adopted the following method. We solved the massive
$TBA$ equation for a left-right symmetric configuration%
\footnote{To verify that the solution is correct one can compare it to numerical
results from TCSA \cite{Yurov:1989yu}.%
}, and decreased $mL$ to a small value (down to $10^{-3}$) to get
a starting position for the pseudo energy $\epsilon$ and the singularity
positions $\bar{\theta}_{j}$ in the massless limit. Then the solution
was refined by iterating the massless TBA system (\ref{eq:massless_lytba}),
and the consistency of the solution verified by matching the value
of $c_{\mathrm{eff}}$ against the value expected from CFT. For the
case of the second, third and fifth excited states it is only necessary
to consider a single pair of singularities $\bar{\theta},\bar{\theta}^{*}$
whose position is

\begin{center}
\begin{tabular}{|c|c|c|}
\hline 
State & $c_{\mathrm{eff}}$ & $\bar{\theta}-i\frac{\pi}{3}$\tabularnewline
\hline 
\hline 
$L_{-1}\varphi$ & $-\frac{118}{5}$ & $1.84285939+0.006848815\, i$\tabularnewline
\hline 
$L_{-1}^{2}\varphi$ & $-\frac{238}{5}$ & $2.53410313+3.03963559\times10^{-5}i$\tabularnewline
\hline 
$L_{-1}^{3}\varphi$ & $-\frac{358}{5}$ & $2.93773630+1.3531641\times10^{-6}i$\tabularnewline
\hline 
\end{tabular}
\par\end{center}

\noindent (up to the accuracy shown).

The case when the infrared (large $mL$) configuration contains a
stationary particle is somewhat exceptional; this is the case for
the first excited state which corresponds to the identity operator.
For large $mL$ the position of the singularity is given by 
\[
\bar{\theta}\sim i\left(\frac{\pi}{6}+\sqrt{3}\mathrm{e}^{-\sqrt{3}mL/2}\right)
\]
but for small $mL$ the imaginary part of the position of the singularities
approaches the values $i\pi/3$ and eventually hits it, after which
one obtains a pair of singularities $i\pi/3\pm\alpha$ of which the
right-moving one survives in the kink limit. Due to (\ref{eq:yeqn})
this leads to a zero of the $Y(\theta)$ at $\theta=\alpha$, which
means that the pseudo-energy function $\epsilon(\theta)$ diverges
to $-\infty$ at the same point. To treat this problem we chose the
prescription given in \cite{BLZ:9607099} and redefined the pseudo-energy
by extracting the singular term:
\[
\bar{\epsilon}_{1}(\theta)=\epsilon_{1}(\theta)-\log\tanh\frac{3}{4}(\theta'-\alpha)
\]
so that the integral equation for the massless limit becomes
\[
\bar{\epsilon}_{1}(\theta)=\mbox{e}^{\theta}-\int_{-\infty}^{\infty}\frac{d\theta'}{2\pi}\varphi(\theta-\theta')\log\left(\tanh\frac{3}{4}(\theta'-\alpha)+\mathrm{e}^{-\bar{\epsilon}_{1}(\theta')}\right)
\]
It is supplemented by the following condition for the singularity
position
\[
\sqrt{3}\mbox{e}^{\alpha}+\frac{3}{\pi}\fint_{-\infty}^{\infty}\frac{d\theta'}{2\pi}\frac{\cosh2(\theta'-\alpha)}{\sinh3(\theta'-\alpha)}\log\left(\tanh\frac{3}{4}(\theta'-\alpha)+\mathrm{e}^{-\bar{\epsilon}_{1}(\theta')}\right)=\pi
\]
where $\fint$ denotes the principal value of the singular integral.
These equations can now be iterated as described above and $Y_{1}$
can be expressed as
\[
Y_{1}(\theta)=\tanh\frac{3}{4}(\theta-\alpha)\mathrm{e}^{\bar{\epsilon}_{1}(\theta)}
\]
The singularity position for this state is 
\[
\alpha=0.49577315\dots
\]
and the value of the effective central charge is
\[
c_{\mathrm{eff}}=\frac{6\sqrt{3}}{\pi}\mathrm{e}^{\alpha}+\frac{6}{\pi^{2}}\int_{-\infty}^{\infty}d\theta\mathrm{e}^{\theta}\log\left|1+Y_{1}(\theta)^{-1}\right|=-\frac{22}{5}
\]
(due to $Y_{1}(\alpha)=0$, the integrand has a logarithmic singularity,
which is integrable, but makes a precise numerical evaluation harder).

For the 4th excited state corresponding to the energy momentum tensor
$T=L_{-2}\mathrm{id}$, $Y_{4}$ has a real zero $\alpha$ together
with an ordinary complex $\bar{\theta}$, so that
\[
Y_{4}(\theta)=\tanh\frac{3}{4}(\theta-\alpha)\mathrm{e}^{\bar{\epsilon}_{4}(\theta)}
\]
where $\bar{\epsilon}_{4}$ satisfies
\[
\bar{\epsilon}_{4}(\theta)=\mbox{e}^{\theta}+\log\frac{S(\theta-\bar{\theta})}{S(\theta-\bar{\theta}^{*})}-\int_{-\infty}^{\infty}\frac{d\theta'}{2\pi}\varphi(\theta-\theta')\log\left(\tanh\frac{3}{4}(\theta'-\alpha)+\mathrm{e}^{-\bar{\epsilon}_{4}(\theta')}\right)
\]
The positions of the singularities can be determined from the equations
\begin{eqnarray*}
\pi & = & \sqrt{3}\mbox{e}^{\alpha}-i\left[\log\frac{S(\alpha+i\pi/3-\bar{\theta})}{S(\alpha+i\pi/3-\bar{\theta}^{*})}-\log\frac{S(\alpha-i\pi/3-\bar{\theta})}{S(\alpha-i\pi/3-\bar{\theta}^{*})}\right]\\
 &  & +\frac{3}{\pi}\fint_{-\infty}^{\infty}\frac{d\theta'}{2\pi}\frac{\cosh2(\theta'-\alpha)}{\sinh3(\theta'-\alpha)}\log\left(\tanh\frac{3}{4}(\theta'-\alpha)+\mathrm{e}^{-\bar{\epsilon}_{4}(\theta')}\right)\\
0 & = & 1+Y_{4}(\bar{\theta})
\end{eqnarray*}
with the results
\[
\alpha=0.83331625\dots\qquad\bar{\theta}=2.49331757\dots+0.52367139\dots\times i
\]
and the effective central charge is
\begin{eqnarray*}
c_{\mathrm{eff}} & = & \frac{6\sqrt{3}}{\pi}\mathrm{e}^{\alpha}+\frac{12i}{\pi}\left(\mathrm{e}^{\bar{\theta}}-\mathrm{e}^{\bar{\theta}^{*}}\right)+\frac{6}{\pi^{2}}\int_{-\infty}^{\infty}d\theta\mathrm{e}^{\theta}\log\left|1+Y_{4}(\theta)^{-1}\right|\\
 & = & -\frac{262}{5}
\end{eqnarray*}

\newcommand\arxiv[2]      {\href{http://arXiv.org/abs/#1}{#2}}
\newcommand\doi[2]        {\href{http://dx.doi.org/#1}{#2}}
\newcommand\httpurl[2]    {\href{http://#1}{#2}}


\newpage
\begin{thebibliography}{XXX}
\setlength{\itemsep}{-0em}
\small

\bibitem{YZ1}
\doi{10.1142/S0217751X9000218X}{%
V.P.~Yurov and A.B.~Zamolodchikov,
{\em Truncated conformal space approach to the scaling Lee-Yang
  model},
Int.~J.~Mod.~Phys {\bf A5} (1990) 3221--3245.
}

\bibitem{Dorey:1997yg}
\doi{10.1016/S0550-3213(98)00339-3}{P.~Dorey, A.~Pocklington, R.~Tateo and G.~Watts,
  {\it TBA and TCSA with boundaries and excited states},
  Nucl.\ Phys.\  B {\bf 525} (1998) 641
  [hep-th/9712197].}

\bibitem{Dorey:2000}
\doi{10.1016/S0550-3213(99)00772-5}{P.~Dorey, I.~Runkel, R.~Tateo and G.~Watts,
  {\it g function flow in perturbed boundary conformal field theories},
  Nucl.\ Phys.\  B {\bf 578} (2000) 85--122
  [hep-th/9909216].}

\bibitem{Zamo:sphere}
\arxiv{hep-th/0109078}{%
Al.~Zamolodchikov,
{\it Scaling Lee-Yang model on a sphere. 1. Partition function},
JHEP 0207:029,2002
[arxiv:hep-th/0109078]}

\bibitem{ishibashi}
\doi{10.1142/S0217732389000320}{N.~Ishibashi,
{\em The Boundary and Crosscap States in Conformal Field Theories},
Mod.~Phys.~Lett.~{\bf A4} (1989) 251}

\bibitem{AffleckLudwig}
\doi{10.1103/PhysRevLett.67.161}{%
I.~Affleck and A.W.W.~Ludwig,
{\em Universal noninteger 'ground state degeneracy' in critical quantum systems},
Phys.~Rev.~Lett.~{\bf 67} (1991) 161--164.}

\bibitem{LMSS}
\doi{10.1016/0550-3213(95)00435-U}{%
A.~LeClair,  G.~Mussardo, H.~Saleur and S.~Skorik,
{\em Boundary energy and boundary states in integrable quantum field theories},
Nucl.~Phys.~{\bf B453} (1995) 581--618
[hep-th/9503227]}

\bibitem{BLZ:9607099}
\doi{10.1016/S0550-3213(97)00022-9}{%
V.V.~Bazhanov, S.L.~Lukyanov and Al.B.~Zamolodchikov,
{\em 
Integrable quantum field theories in finite volume: excited state energies},
Nucl.~Phys.~B {\bf 489} (1997) 487--531
[arxiv:hep-th/9607099]}

\blank{\bibitem{Bazhanov:1996aq}
V.~V. Bazhanov, S.~L. Lukyanov and A.~B. Zamolodchikov, 
{\em Quantum field
  theories in finite volume: Excited state energies,}
Nucl.~Phys.~{\bfseries B489} (1997) 487--531,
[hep-th/9607099]
}

\bibitem{Runkel-T}
\doi{10.1088/1751-8113/41/10/105401}{%
I.~Runkel, {\em  Perturbed Defects and T-Systems in Conformal Field Theory},
J.~Phys.~{\bf A41} (2008) 105401,
[arXiv:0711.0102 [hep-th]]}

\bibitem{GW:BTCSARG}
\arxiv{arXiv:1104.0225}
{G.M.T.~Watts,
{\em On the renormalisation group for the boundary Truncated Conformal
  Space Approach},
KCL-MTH-11-04, 
[arXiv:1104.0225]
}

\bibitem{Wynn}
{P.~Wynn,
{\em On a Device for Computing the $\epsilon_m(S_n)$ transformation},
Math Tables Aids Comput. 10 
(1956) 91-96
}

\providecommand{\href}[2]{#2}

\bibitem{Zamolodchikov:1989cf}
\doi{10.1016/0550-3213(90)90333-9}{%
A.~B. Zamolodchikov, 
{\em Thermodynamic Bethe Ansatz in relativistic models.
  scaling three state Potts and Lee-Yang models,}
Nucl.~Phys.~{\bfseries B342} (1990) 695--720.
}

\bibitem{Dorey:1996re}
\doi{10.1016/S0550-3213(96)00516-0}{%
P.~Dorey and R.~Tateo, 
{\em Excited states by analytic continuation of TBA
  equations,}
Nucl.~Phys.~{\bfseries B482} (1996) 639--659,
[hep-th/9607167]
}


\bibitem{Zamolodchikov:1991et}
\doi{10.1016/0370-2693(91)91737-G}{%
A.~B. Zamolodchikov, {\em On the thermodynamic Bethe ansatz equations for
  reflectionless ADE scattering theories},
{Phys.~Lett.~}{\bfseries B253} (1991) 391--394.
}

\bibitem{Yurov:1989yu}
\doi{10.1142/S0217751X9000218X}{%
V.~P. Yurov and A.~B. Zamolodchikov, 
{\em Truncated conformal space approach to
  scaling Lee-Yang model,}
{Int.~J.~Mod.~Phys.}~{\bfseries A5} (1990) 3221--3246.
}

\bibitem{DT}
\doi{10.1016/j.nuclphysb.2004.06.045}{%
P.E.~Dorey, D.~Fioravanti, C.~Rim and R.~Tateo,
{\em Integrable quantum field theory with boundaries: The Exact g function},
Nucl. Phys. {\bf B696} (2004) 445--467,
[hep-th/0404014]}

\bibitem{P}
\doi{10.1007/JHEP08(2010)090}{%
B.~Pozsgay,
{\em  On O(1) contributions to the free energy in Bethe Ansatz systems: The Exact g-function},
JHEP 1008:090,2010.
[arXiv:1003.5542 [hep-th]]}

\bibitem{quella}
\doi{10.1088/1126-6708/2007/04/095}{%
T.~Quella, I.~Runkel and G.M.T.~Watts,
{\em  Reflection and transmission for conformal defects,}
JHEP 0704:095,2007.
[hep-th/0611296]}

\bibitem{marton}
\doi{10.1088/1126-6708/2009/11/057}{%
M.~Kormos, I.~Runkel and G.M.T.~Watts,
{\em Defect flows in minimal models,}
JHEP 0911:057,2009
[arXiv:0907.1497 [hep-th]]
}

\bibitem{mathematica}
Wolfram Research Inc, {\em Mathematica}, version 8.0, Champaign
Illinois, 2010.


\end{thebibliography}
\end{document}